\documentclass[12pt]{iopart}

\usepackage{color}
\usepackage{graphicx}
\usepackage{appendix}
\usepackage{bm}
\usepackage{txfonts}
\begin{document}

\title[Monte Carlo calculations of pair production in high-power laser-plasma interactions]{Monte Carlo 
calculations of pair production in high-intensity laser-plasma interactions}

\author{R Duclous$^1$, J G Kirk$^2$ \& A R Bell$^{1,3}$ }

\address{$^1$ Clarendon Laboratory, University of Oxford, Parks Road, Oxford UK OX1 3PU}
\address{$^2$ Max-Planck-Institut f\"ur Kernphysik, Postfach 10 39 80, 69029 Heidelberg, Germany}
\address{$^3$ STFC Rutherford Appleton Laboratory, Didcot, Oxfordshire UK OX11 0QX}
\ead{r.duclous1@physics.ox.ac.uk}

\begin{abstract}

Gamma-ray and electron-positron pair production will figure
prominently in laser-plasma experiments with next generation lasers.
Using a Monte Carlo approach we show that {\em straggling} effects 
arising from the finite recoil an electron experiences when it emits
a high energy photon, increase the 
number of pairs produced on further interaction with the laser fields.
\end{abstract}

\section{Introduction}
\label{intro}
Next generation lasers such as ELI \cite{ELI1,ELI2} and VULCAN
\cite{Vulcan1,Vulcan2} are expected to achieve intensities of
$10^{23} \,\mbox{W cm}^{-2}$ or greater, opening up rich opportunities to study 
QED processes in the strong field regime\cite{salaminetal06,muelleretal09,dipiazzaetal09}. 
One of these processes is pair production in 
counter-propagating beams, which 
is a significant effect at $10^{23}\mbox{W cm}^{-2}$\cite{Bell}, and,
at $10^{24} \,\mbox{W cm}^{-2}$ is sufficiently strong that a 
cascade may develop in which each electron
produces a pair that in turn produces further pairs, to initiate an
avalanche. 

In intense laser beams, the dominant route to pair production is a two-stage
process.  Firstly, an electron with Lorentz factor $\gamma$ interacts
with the laser fields ($\bm{E}$,$\bm{B}$) to produce a gamma-ray with
high energy $h\nu $.  Secondly, the gamma-ray interacts with the
same fields as it propagates to produce an electron-positron
pair \cite{Bell,PPCF}.  The crucial parameters for the two-stage process
are (i) $\eta \approx \gamma |{\bm E}_\perp+{\bm v}\times {\bm B}|
/E_{\rm crit}$\footnote{This expression is generally 
accurate for $\gamma\gg1$, a precise definition is given in (\ref{etachidef})}
where $E_{\rm crit}=1.3\times 10^{18}\,{\rm Vm}^{-1}$ is the
Schwinger field, and (ii) $\chi = h\nu |{\bm E}_\perp+(c{\bm k}/k)\times
{\bm B}|/2m_{\rm e}c^2 E_{\rm crit}$ where $\hbar\bm{k}$ is the gamma-ray
momentum and $\bm{E}_\perp$ is the component of the electric field 
perpendicular to the
direction of motion of the electron or photon, as appropriate.
Strong field QED effects such as pair production
become important when these parameters reach unity.
The energy of the gamma-ray photon is typically $h\nu\approx0.44\eta \gamma
mc^2$, implying $\chi\approx 0.22 \eta^2$, 
and the cross-section for pair production is proportional to
$\exp (-4/3\chi)$ for $\chi\ll 1$.  Because of the exponential cut-off
to the cross-section, the process is very sensitive to the value 
of $\eta$ reached by the
electron, when the regime of strong QED is approached.  
This leads to an abrupt laser-intensity threshold for pair
production as noted by \cite{Bell}, \cite{PPCF} and
\cite{fedotov10}, where a small increase
in $\eta$ can lead to a large increase in pair production. 

In \cite{Bell}, \cite{PPCF} and \cite{Sokolov1}, a semi-classical
treatment was used in which electrons were subject to a continuous
loss of energy through radiation of gamma-rays. In this continuous
model, the radiation reaction force is approximated, to the lowest
order in $1/\gamma$, by the expression
\begin{eqnarray}
\label{fradg}
\displaystyle {\bf f}_{rad}  =   -\frac{2}{3} \alpha_f \frac{m_{\rm e}^2c^3}{\hbar}\eta^2g(\eta) \frac{{\bf p}}{p} 
\quad , \quad g(\eta) = \frac{3\sqrt{3}}{2\pi \eta^2} \int_0^{\infty}F(\eta,\chi)d\chi \quad ,
\end{eqnarray} 
where $g(\eta) \in [0,1]$ accounts for the reduction of the total
power radiated by the electron. $F(\eta,\chi)$ is the quantum
synchrotron emissivity, as defined in Appendix~A of
\cite{PPCF}.  However, in reality,
gamma-ray emission is a random quantum process \cite{Shen1,Shen2}.  As
$\eta$ approaches unity, the energy of the emitted gamma-ray becomes a
substantial fraction, typically $0.44\eta$, of the kinetic energy of
the electron before emission, and the resulting electron trajectory
starts to fluctuate substantially away from the semi-classical
average.  This enhances pair production in two ways. Firstly, if an
accelerating electron propagates by chance over an unusually large
distance before emitting a gamma-ray, it may reach a large Lorentz
factor $\gamma$ and emit a gamma-ray with an unusually high
energy. This, in turn, has a much increased probability of producing a
pair, as compared to the photons emitted by an electron moving on a
classical trajectory in the same field. Such events cumulate, and we
show in \ref{Strag} that this straggling effect \cite{Shen1} enhances
pair production.  Secondly, the deviations in the trajectory caused by
straggling lead the electron to sample the laser fields at locations
other than those on the classical path. In \cite{PPCF}, it was shown
that, under the influence of continuous radiation losses, electrons
naturally migrate towards points in the laser field at which they emit
few high-energy gamma-rays. In the case of counter-propagating
circularly polarised waves, these are the nodes in the electric field.
Once an electron settles at an $\bm{E}=0$ node, its Lorentz factor
drops and pair production ceases.  In contrast, we show in
\ref{offnode} that straggling causes electrons to migrate more slowly
towards these nodes. Consequently, electrons undergoing discontinuous
momentum changes retain a large Lorentz factor for longer than their
semi-classical counterparts, and emit a larger number of high energy
gamma-rays, which subsequently produce more pairs.

In this paper, we extend the calculations of \cite{Bell} and
\cite{PPCF} by modelling radiation loss as a discontinuous process
that leads to straggling effects on the electron trajectory. 
This is done using a Monte-Carlo method that is described in 
section~\ref{MCalgo}. 
In section
\ref{comparMC}, we compare the continuous with the more realistic 
discontinuous loss case,
and show that the latter produces a
larger number of pairs at laser intensities around $10^{23} \,\mbox{W
  cm}^{-2}$.  Pair 
production at $10^{24} \,\mbox{W cm}^{-2}$ is 
relatively little affected by straggling since, at this intensity, 
even semi-classical, continuous electron trajectories lead to 
gamma-rays that are 
well over the threshold for pair production.


\section{Description of the Monte Carlo algorithm}
\label{MCalgo}
In 4-vector
notation, the quantities $\eta$ and $\chi$ are defined as
\begin{eqnarray}
\label{etachidef}
\eta&=& \frac{e\hbar}{m_{\rm e}^3c^4} \left |F_{\mu \nu}p^{\nu} \right | 
\quad , \quad
\chi\,=\, \frac{e\hbar^2}{2m_{\rm e}^3c^4} \left | F_{\mu
  \nu}k^{\nu} \right |,
\end{eqnarray} 
for an electron and photon whose four-momenta are
$p^{\nu}$ and $\hbar k^{\nu}$, respectively, with
$ F_{\mu \nu}$ the electromagnetic field tensor. In the
following, we make use of the approach presented in \cite{PPCF},
that retains the weak, quasi-stationary field approximation. The
probabilities for gamma-ray photon
emission and the subsequent pair production are in this approximation 
functions of
$\eta$ and $\chi$ only.  An equivalent system is then chosen, for the same
given $\chi$ and $\eta$, in which the electron moves in a plane
perpendicular to a uniform, static $B$ field.  This allows us to use
the transition probabilities as formulated in \cite{Erber}.

The radiation emitted by the particle due to its acceleration in the 
laser fields --- here called synchrotron radiation --- 
is assumed to be a random walk process
\cite{Shen1,Shen2}. The probability of emission of a gamma-ray photon is
governed by an optical depth $\tau_{\rm e}$.  At the start of the
calculation, and immediately following the emission of a
photon, the current optical depth $\tau_{\rm e}$ is set to zero and a
randomly chosen \lq final\rq\ optical depth $\tau_{\rm e}^{(\rm f)}$ is assigned to the
electron.  The current optical depth $\tau_{\rm e}$ increases as the
electron propagates until it reaches $\tau_{\rm e}^{(\rm f)}$, at which point a
photon is emitted.  At each computational timestep the electron's
position, momentum and current optical depth are updated according to 
the equations
\begin{eqnarray}
\label{MC1}
&& \displaystyle \frac{\mbox{d}{\bm p}}{\mbox{d} t} = - e \left ({\bm
  E}+{\bm v} \wedge {\bm B} \right ) \ , \\\label{MC2} &&
\displaystyle \frac{\mbox{d}{\bm x}}{\mbox{d} t} = \frac{\bm
  p}{m_{\rm e}\gamma} \ , \\\label{MC3} && \displaystyle
\frac{\mbox{d}\tau_{\rm e}}{\mbox{d} t} =
\int_0^{\eta/2}\frac{\mbox{d}^2N}{\mbox{d}\chi \mbox{d}t} (\eta,\chi)
\,\mbox{d} \chi \ ,
\end{eqnarray}
where
\begin{eqnarray}
 \displaystyle\frac{\mbox{d}^2N}{\mbox{d}\chi \mbox{d}t}
(\eta,\chi) &=& \frac{\sqrt{3}}{2\pi \tau_{\rm C}} \alpha_{\rm f}
\frac{\eta}{\gamma} \frac{F(\eta,\chi)}{\chi}
\end{eqnarray}
is the differential rate of production of photons of parameter $\chi$
by an electron of parameter $\eta$ and
$F(\eta,\chi)$ is the quantum synchrotron emissivity, expressed in
the weak, quasi-stationary field approximation \cite{Erber}. An
expression for $F(\eta,\chi)$ is given in Appendix~A of
\cite{PPCF}. The Compton time is $\tau_{\rm C}=\lambdabar_{\rm C}/c$, where 
 $\lambdabar_{\rm C} = \hbar /m_{\rm e}c $ is the Compton
wavelength, and $\alpha_{\rm f}$ is the fine structure constant.

\begin{figure}[htbp]
\begin{center}
\includegraphics[width=8.cm,height=7.cm,angle=0]{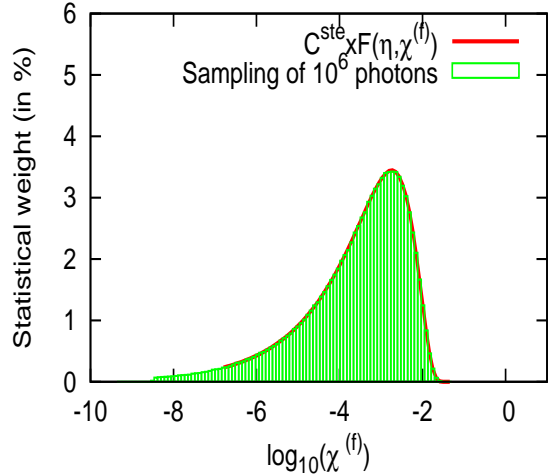}
\caption{Comparison between the distribution in $\chi^{(f)}$
 resulting from the sampling relation (\ref{PHenergySample})
  and the quantum synchrotron emissivity $F(\eta,\chi^{(f)})$.}
\label{convF}
\end{center}
\end{figure}

When $\tau_{\rm e}=\tau_{\rm e}^{(\rm f)}$ is reached, a photon is emitted, and its
 energy is randomly assigned as follows.  First, the parameter $\chi ^{(\rm f)}$ is 
found from the relation
\begin{eqnarray}
\label{PHenergySample}
&&\displaystyle \xi = \frac{\int_0^{\chi^{(\rm f)}}
 \frac{\mbox{d}^2N}{\mbox{d}\chi \mbox{d}t}  \mbox{d} \chi  }{\int_0^{\eta/2}
 \frac{\mbox{d}^2N}{\mbox{d}\chi \mbox{d}t}  \mbox{d} \chi } \quad  \left (  = \frac{\int_0^{\chi^{(\rm f)}}
 \frac{F(\eta,\chi)}{\chi} \mbox{d} \chi  }{\int_0^{\eta/2}
 \frac{F(\eta,\chi)}{\chi} \mbox{d} \chi } \right ) \ ,
\end{eqnarray}
where $\xi $ is a uniformly distributed random number in [0,1], and
the right hand side is a monotonic, increasing function of
$\chi^{(\rm f)}$, which is tabulated in both the $\chi^{(\rm f)}$ and $\eta$
directions. The variables $\chi^{(\rm f)}$ and $\eta$, in equation
(\ref{PHenergySample}), depend on the electromagnetic field at the
point of photon emission. 
Then, assuming that the photon is emitted parallel to the electron 
momentum, its energy $h\nu$, is determined by the relation
\begin{eqnarray}
\displaystyle {h\nu} &=& \frac{2m_{\rm e}c^2\chi^{(\rm f)}\gamma}{\eta} \ . 
\end{eqnarray}
which follows from the definitions of $\eta$ and $\chi$ and the
assumption that the electron is relativistic, $\gamma\gg1$.  For
computational efficiency, we ignore the rare photons emitted in the
low energy part of the spectrum, $\chi^{(\rm f)}<\chi^{(\rm f)}_{\rm
  min}$, where $\chi^{(\rm f)}_{\rm min}$ is chosen such that the
neglected photons carry only $10^{-9}$ of the total energy and are
incapable of producing pairs. The computational procedure was
successfully tested by its ability to reproduce the correct
synchrotron spectrum. This is shown in Fig.\ref{convF}, for a large
number ($10^6$) of emitted photons in a regime where $\eta=0.1$. A
$\chi^{(f)}$ parameter, which represents the photon energy, is
assigned to each photon at the emission point, according to the
sampling relation (\ref{PHenergySample}). The occurrence (in percent of the
total number of the $10^6$ emission events) of emissions in each energy interval
gives a distribution along the $\chi^{(f)}$ axis, that scales with the
quantum synchrotron emissivity $F(\eta,\chi^{(f)})$.

After the electron has emitted a photon, which is assumed to occur 
instantaneously, it continues its
trajectory beginning at the same point but with a different 
energy $\gamma^{(\rm f)}mc^2$ and momentum ${\bm p}^{(\rm f)}$.
To find $\bm{p}^{(\rm f)}$, we again
use the
approximation that the photon is emitted parallel to the
pre-emission electron momentum, so that, by conservation of momentum, 
\begin{eqnarray}
\bm{p}^{(\rm f)}&=&\left(1- \frac{h\nu}{cp}\right)\bm{p}\,.
\label{pchange}
\end{eqnarray}
In reality, the photon may be emitted over a
cone of opening angle approximately $1/\gamma$, but we neglect this 
small angular spread here. Equation (\ref{pchange}) implies
that a small amount of energy is extracted from the
laser fields during photon emission, which we also neglect. 
This neglect is unimportant in the 
present test-particle model in which the laser fields are prescribed, since
the fraction of the electron energy transferred to the laser
field is small $\sim 1/(\gamma \gamma^{(\rm f)})$, as shown in
\ref{AppendixA}. However this point may need more careful treatment in a more
complete dynamic treatment of the development of an electron-positron
cascade and its inclusion in a particle-in-cell code
\cite{SokolovQED,Timokhin}.


\section{Straggling effects}
\label{comparMC}

As in \cite{PPCF}, we adopt the configuration of two
counter-streaming, circularly polarized laser beams, which can
qualitatively be considered equivalent to the interaction of an
incident and reflected beam from a solid \cite{Bell}. We restrict the
treatment to pair creation by the two-step process described in
section \ref{MCalgo}.  Pairs can also be produced directly by an electron
interacting with the laser field, without the need to produce an
intermediate real photon. This process is analogous to trident pair
production by an electron in the Coulomb field of a nucleus. However,
although this direct, single-step process may be important at lower
laser intensity, it is dominated by the indirect two-step process at
laser intensities above $10^{23} \,\mbox{Wcm}^{-2}$, and we do not
include it in the results presented here.

 In the two-step process, the emitted gamma-rays travel through the
 plasma and may produce pairs according to the cross-section which
 depends on $\chi$.  In many ways, the process is comparable with that
 of gamma-ray production by electrons. The photon trajectory could be
 followed through the electromagnetic fields and the optical depth
 integrated as for electrons.  However, in order to make a direct
 comparison with the rate computed assuming continuous losses, we here
 follow the much simpler approximate approach that was used in
 \cite{PPCF}. The probability of pair production is calculated for a
 gamma-ray with $\chi=\chi^{(\rm f)}$, that propagates through
 electromagnetic fields that are assumed to equal those at the point
 where the gamma-ray was emitted.  We follow \cite{PPCF} and
 make the assumption that conversion into pairs can occur until the
 photon has propagated a distance of one laser wavelength.  Here the
 approximation is that the beams have a transverse structure with
 cylindrical geometry, whose radius is one wavelength, representative
 of beams which are tightly focussed at the diffraction limit to
 achieve maximum laser intensity.  The real photons only produce pairs
 while they propagate through the region of large laser field. In
 contrast to electrons, whose trajectory is determined by the
 electromagnetic fields, the photons are undeflected. This means that
 all along its path through the laser fields, the value of the
 parameter $\chi$ remains unchanged at $\chi=\chi^{(\rm f)}$.  Further
 details can be found in section 3.3 of \cite{PPCF}. The differential
 pair production rate {\it via} real photons is the product of the
 differential rate for real photon emission, $
 \displaystyle\frac{\mbox{d}^2N}{\mbox{d}\chi \mbox{d}t} (\eta,\chi)$,
 and an effective conversion probability into pairs, $\displaystyle
 \left ( 1 - \exp ( - \left < \tau_{\rm p} \right > )\right ) $, where
 $\displaystyle \left < \tau_{\rm p} \right > $ is an effective
 optical depth to pair conversion. The rate of pair production {\it
   via} real photons per electron is then addressed with the same rate
 equation as in \cite{PPCF}
\begin{eqnarray}
\label{rate_equation}
\displaystyle \frac{\mbox{d}N_{\rm real}}{\mbox{dt}} &=&
\int_0^{\eta/2}\mbox{d}\chi \frac{\mbox{d}^2N}{\mbox{d}\chi
  \mbox{d}t}(\eta,\chi) \left ( 1 - \exp ( - \left < \tau_{\rm p} \right > )
\right ) \quad .
\end{eqnarray}

\subsection{Cumulative effects of the straggling at a $B=0$ node}
\label{Strag}

In this section, we select a particular trajectory at a $\bm{B}=0$ node
(zero magnetic field) of an infinite standing wave generated by two
counter-propagating, circularly polarized laser beams with the same
sense of rotation of the fields. Throughout, distances and times are
scaled with the laser wavelength and laser period,
respectively. This configuration has been extensively analysed in
\cite{Bell,PPCF} in the case of continuous radiation losses where it
was shown that the electron settles into a circular orbit, which is
favourable for pair production. Here we show how discontinuous
trajectories increase pair production.

\begin{figure}[htbp]
\begin{center}
\begin{tabular}{cc}
\includegraphics[width=7.5cm,height=7.cm,angle=0]{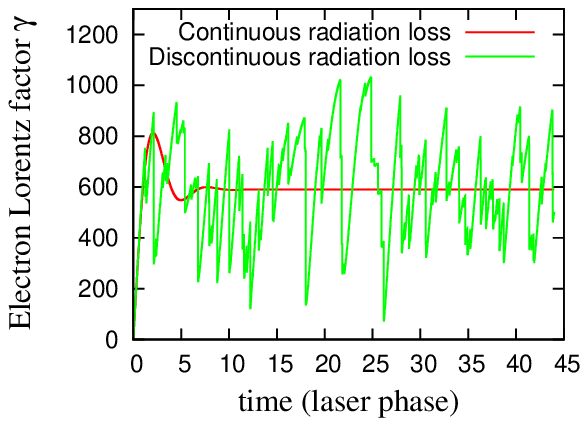} &
\includegraphics[width=7.5cm,height=7.cm,angle=0]{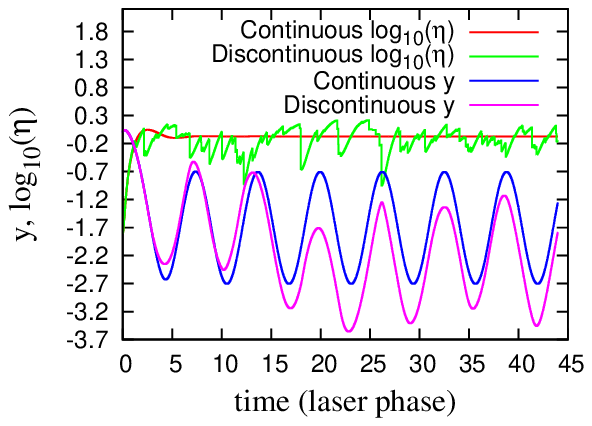}
\\ $(a)$ & $(b)$
\end{tabular}
\caption{Case of an electron at a $\bm{B}=0$ node of two very long,
    counter-propagating, circularly polarized pulses. The trajectories found 
with continuous and discontinuous loss-descriptions are compared in terms
    of (a) Lorentz factor, (b) $\eta$-parameter \& spatial 
coordinate $y$.}
\label{structure}
\end{center}
\end{figure}

\begin{figure}[htbp]
\begin{center}
\includegraphics[width=8.cm,height=7.cm,angle=0]{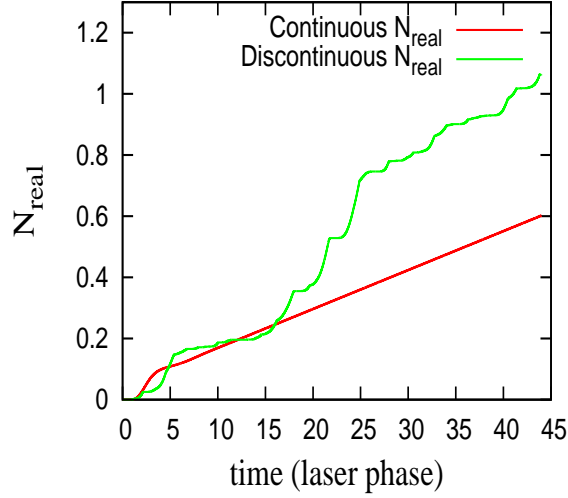}
\caption{Number of pairs produced against time, by the continuous-loss
trajectory (red curve) and the discontinuous one (green curve).}
\label{pairs}
\end{center}
\end{figure}

The two beams counter-propagate along the $z$-axis, each with an
intensity $I_{24} \times 10^{24} \,\mbox{W cm}^{-2}$, with
$I_{24}=0.3$. The electron is initially located at a $\bm{B}=0$ node
in the standing wave formed by the counter-propagating beams. Its
initial Lorentz factor is $\gamma=10$, with its momentum oriented
along the positive $y$-axis.  The electron motion is tracked over
seven laser periods during which it remains at the magnetic node.  In
Fig.~\ref{structure}(a), the discontinuous Monte Carlo calculations
lead to the straggling of the Lorentz factor for the particle (green
curve). This is compared with the Lorentz factor of a particle that
undergoes continuous radiation loss, computed as in \cite{PPCF} (red
curve) for the same initial conditions.  A feature of the straggling
in this regime is a $\sim 30\%$ fluctuation about the Lorentz factor
derived from the continuous description. The particle motion, as shown
in Fig.~\ref{structure}(b) is characterized by excursions to higher
and lower $\eta$ values than those reached by an electron undergoing
continuous losses. The discrepancies in the particle $y$ position,
between the two descriptions, is also shown in
Fig.~\ref{structure}(b). Finally, the comparison of Fig.~\ref{pairs},
and Fig.~\ref{structure}(b), shows that the high $\eta$ excursions are
very effective at producing pairs. Moreover, the accumulation of
pairs, produced by the high $\eta$ excursions, leads to an enhancement
of the overall pair production, in this example by about $40\%$ after
only seven laser periods. At this time, the average number of pairs
produces by an electron reaches unity, which can be considered an
approximate threshold for the initiation of an electron-positron
avalanche.

\subsection{Convergence properties of discontinuous trajectories, off-$B=0$ nodes}
\label{offnode}
In \cite{PPCF}, the authors showed that the electron trajectory at a
$\bm{B}=0$ node is unstable. Small displacements from the node grow, causing
the electron to migrate towards the stable $\bm{E}=0$ nodes. When the
electron reaches an $\bm{E}=0$ node, its oscillatory momentum damps towards
zero and pair production ceases.  Here, we show that in the more
realistic Monte Carlo description, the migration progresses less
smoothly, and, on average, the electrons tend to take longer to
reach to the $\bm{E}=0$ node.  This increases the number of pairs produced,
since, on average, electrons emit high energy photons for a longer period of
time.

\begin{figure}[htbp]
\begin{center}
\begin{tabular}{cc}
\includegraphics[width=8.cm,height=7.cm,angle=0]{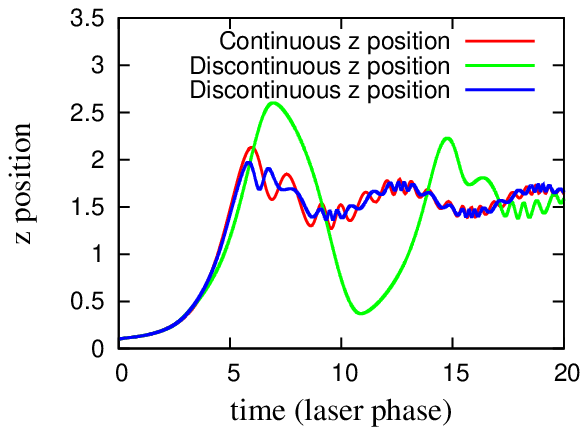}
&
\includegraphics[width=8.cm,height=7cm,angle=0]{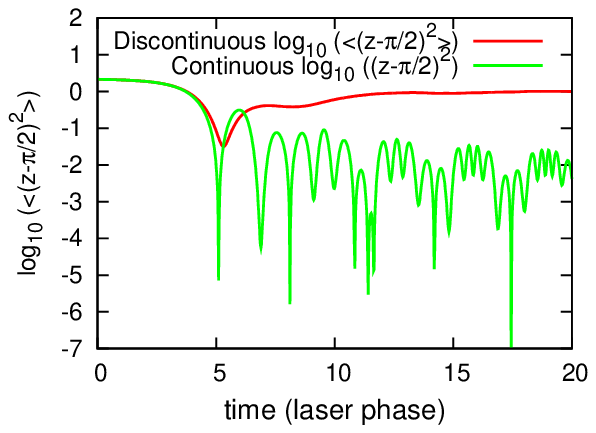}
\\
$(a)$ & $(b)$
\\
 \includegraphics[width=8cm,height=7cm,angle=0]{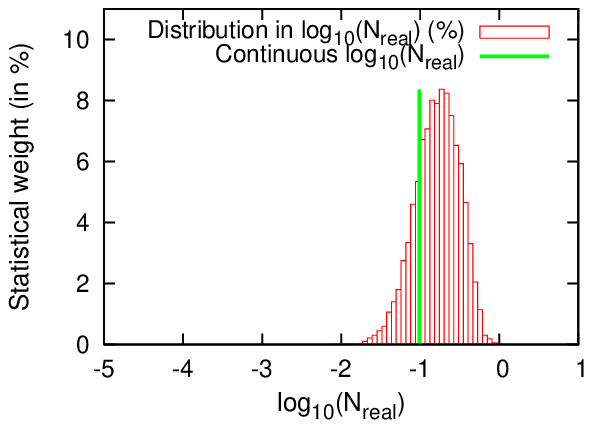}
\\
 $(c)$\\
\end{tabular}
\caption{The discontinuous and continuous descriptions are
    compared at a laser intensity $I_{24}=0.3$, for initialization
    at  $z=0.1$, away from the $\bm{B}=0$ node (located at $z=0$). In
(a) $z$ is shown as a function of time, in (b) the square
    of the distance to the stable $\bm{E}=0$ node (located at $z=\pi/2$), 
and in (c) the statistical 
distribution of the number of pairs $N_{\rm real}$ produced by a trajectory.}
\label{stat}
\end{center}
\end{figure}

We consider the same configuration and initial conditions used in
section \ref{Strag}, with $I_{24}=0.3$, except that the initial
electron position is now $z=0.1$ instead of $z=0$, which is the
location of the $\bm{B}=0$ node.  Then the motion of a set of $10^4$
electrons is followed until the time $t_{\rm max}=20$. Because of
random fluctuations, each of the $10^4$ trajectories is different,
resulting in a different number of pairs finally produced. In
Fig.~\ref{stat}(a), the $z$-position for two sample trajectories is
tracked, and compared with the $z$-position of an equivalent electron
starting from the same point but subject to a continuous
radiation-reaction force. Two differing kinds of behaviour are
highlighted. In one case (the blue curve), an electron trajectory
described by discontinuous radiation losses reaches the stable
$\bm{E}=0$ node earlier than the trajectory described by continuous
radiation losses (red curve). In another discontinuous-loss case
(green curve), the electron takes longer to converge on the stable
$\bm{E}=0$ node. The majority of electrons belong to this latter
class. This is demonstrated quantitatively in Fig.~\ref{stat}(b),
where the red curve shows the evolution of the square of the distance
to the $\bm{E}=0$ node at $z=\pi/2$, averaged over $10^3$
discontinuous trajectories, i.e., $\left < \left ( z- \pi/2 \right )^2
\right >$. For comparison, the green curve shows the case of
continuous radiation loss. The longer migration time, together with
the excursions to large $\eta$ discussed above, combine to enhance
pair production even further.  The overall increase in pair production
is shown in Fig.~\ref{stat}(c) which presents the distribution
function for pair production compiled from $10^4$ trajectories. The
$N_{\rm real}$ axis is divided into $100$ equally spaced intervals in
the logarithmic range $\log_{10}(N_{\rm real}) \in [-5,1]$. All the
electrons are identically initialized at $z=0.1$, with $I_{24}=0.3$
(red bars). We find that the spectrum is substantially shifted upwards
with respect to the continuous value (green line), $\log_{10}(N_{\rm
  real}) \simeq -1.01$, obtained with continuous radiation loss. The
figure shows that $77.5\%$ of the particles produce more pairs than if
their trajectory was computed in a continuous manner.  A small number
of the sampled trajectories come close to producing one pair per
electron ($\log_{10}(N_{\rm real}) \simeq 0$).

\begin{figure}[htbp]
\begin{tabular}{cc}
\includegraphics[width=8.cm,height=7.cm,angle=0]{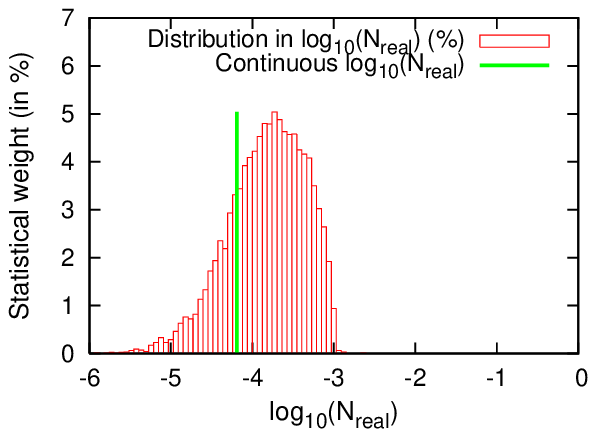}
&
\includegraphics[width=8.cm,height=7.cm,angle=0]{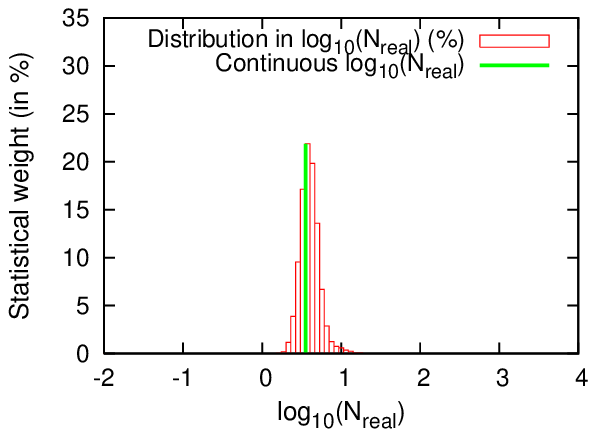}
\\
$(a)$ & $(b)$
\end{tabular}
\caption{The statistical distribution (red bars) of a set of $10^4$ electrons
    identically initialized at $z=0.1$ (close to the $\bm{B}=0$ node
at $z=0$) for
    (a) $I_{24}=0.1$, and (b) $I_{24}=1$. The number of pairs produced
    by an electron undergoing continuous losses is
    shown for comparison (green lines).}
\label{stat_low}
\end{figure}

For the same configuration and initial conditions, we now repeat the
calculation at a lower laser intensity, $I_{24}=0.1$. The results are
presented in Fig.~\ref{stat_low}(a), where the $N_{\rm real}$ axis is
again divided into $100$ equally spaced intervals, this time in the
logarithmic range $\log_{10}(N_{\rm real}) \in [-6,0]$. We find that
the distribution is again shifted upwards with respect to the value
$\log_{10}(N_{\rm real}) \simeq -4.19$ obtained with the continuous
radiation loss. Electron energy straggling and longer migration times
prove to have a significant impact on the number of pairs produced.
For the top $1\%$ of the trajectories, one order of magnitude more
pairs are produced than are predicted using continuous radiation
losses.

In contrast, at higher laser intensities 
the spread of the distribution is quite narrow, and is centred close to 
the continuous value. This is illustrated in Fig.~\ref{stat_low}(b),
where we present results for $I_{24}=1$.  
In this figure, 
the $N_{\rm real}$ axis is divided into $100$ equally spaced intervals in the
logarithmic range $\log_{10}(N_{\rm real}) \in [-2,4]$. The location of
the peak in the spectrum agrees well with the number of pairs computed
with a continuous semi-classical description
$\log_{10}(N_{\rm real}) \simeq 0.55$.
At high laser intensity, straggling has less
effect, because the process is above
threshold and less dependent on statistically rare excursions to large
$\eta$. 

In order to obtain a more straightforward figure of merit, in terms of
produced pairs, we compute an effective value for the number of pairs
produced per electron whose motion is followed:
$$ \displaystyle \left < N_{\rm real}\right > = \sum_{i\in I}^{}
N_{\rm real}(i)P(i) \quad , $$ where $I$ stands for the complete set of
equally spaced intervals that discretize the $N_{\rm real}$ axis. The
index $i$ refers to one of these intervals, in which the probability
is $P(i)$.  This procedure amounts to summing over histograms such as
those in Fig. \ref{stat}(c), \ref{stat_low}(a) \& \ref{stat_low}(b),
for the total expected number of pairs produced per electron starting
from an initial position $z=0.1$.

\begin{figure}[htbp]
\begin{tabular}{cc}
\includegraphics[width=8.cm,height=7.cm,angle=0]{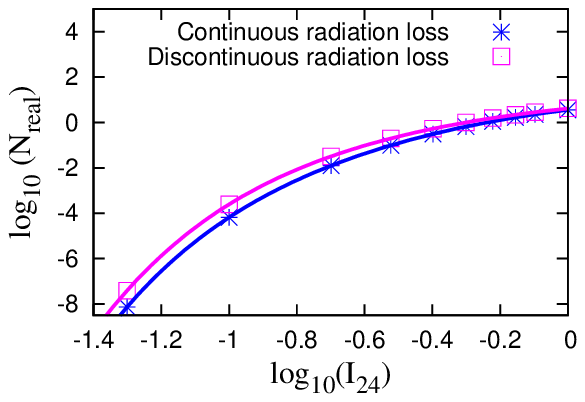}
&
\includegraphics[width=8.cm,height=7.cm,angle=0]{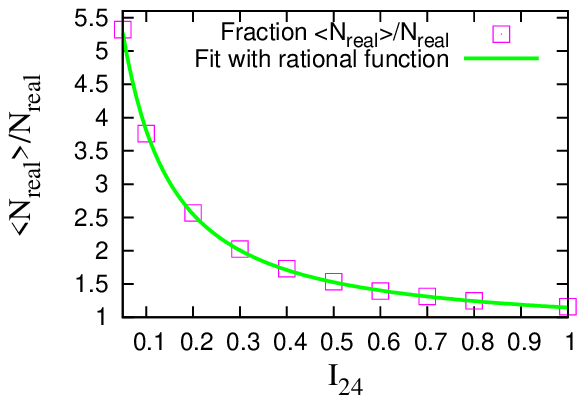}
\\
$(a)$ & $(b)$
\end{tabular}

\caption{The continuous- and discontinuous-loss descriptions are
    compared in terms of the number of pairs produced, with respect to the
    laser intensity parameter $I_{24}$, for two counter-streaming,
    circularly polarized pulses, the particle being initialized
    at $z=0.1$, close to the $\bm{B}=0$ node at $z=0$. 
The curves are fits with rational functions.}
\label{N_vs_real}
\end{figure}

For a range of laser intensities,  Fig.~\ref{N_vs_real}(a) shows a 
comparison between the number of pairs produced per electron undergoing 
discontinuous losses (computed using a Monte Carlo approach), 
with the number found for continuous losses. 
For low laser energy, $I_{24}=0.1$, the number of pairs
 $\displaystyle \left < N_{\rm real} \right > $ associated with 
discontinuous-loss trajectories is about $5$ 
times larger
 than the number associated with continuous-loss trajectories. 
This is due to the
 importance of stragglers in this regime, and has a strong dependence on
 the $\eta$-parameter. Because of the large range in $N_{\rm real}$
(eight orders of magnitude) this effect is more easily seen
in Fig~\ref{N_vs_real}(b), which plots the ratio between these two numbers,
 showing that $\displaystyle \left < N_{\rm real} \right > / N_{\rm real}$
shrinks as the laser intensity increases, until it becomes very close to 
unity, at $I_{24}=1$. The strength of the straggling process is found to be
 very sensitive to the laser intensity for about $I_{24}<0.4$.

\section{Conclusion}

In this paper, we report on calculations of the number of electron
positron pairs produced as an electron moves in the field of a laser
with intensity in the range range $ 10^{23}-10^{24}\,\mbox{W cm}^{-2}$.  
In this regime, the dominant route to pair
production involves the emission of gamma-ray photons {\it via}
synchrotron radiation, and their further conversion into
electron-positron pairs \cite{Bell,PPCF}.  
The new aspect of our treatment is that we
take account of the stochastic nature of gamma-ray production, which
leads to discontinuous quantum jumps in the electron energy; using a
Monte-Carlo algorithm we follow the evolution of a large number of
stochastic trajectories that result from each individual set of
initial conditions.

Pair production has been shown to be important in configurations with 
counter-propagating laser beams, that can be generalized to apply to 
laser-solid interactions \cite{Bell,PPCF}. However, these authors 
used a continuous approximation for the radiation reaction force, 
and examined, therefore, one classical trajectory for each
set of initial conditions. 
In order to make a direct comparison 
with our calculations, we investigate a specific, highly idealized 
configuration of the laser fields that was also used as a test-case
by \cite{Bell,PPCF}: that of 
two long, counterpropagating, circularly polarized laser pulses.

Discontinuous, stochastic jumps in the electron energy give rise to a
spread in energy as a function of time known as {\em straggling}.  Our
main result is that straggling increases the number of pairs produced
for given initial conditions, and for a given laser intensity.  This
behaviour is more pronounced for laser intensities less than $
0.4\times 10^{24}\,\mbox{Wcm}^{-2}$ where the transition probabilities
are strongly dependent on $\eta$ and $\chi$; in this range we observe
up to a fivefold increase in the number of pairs produced.  Straggling
effects are less important at higher intensities, where most
gamma-rays are above threshold for pair production and the number of
electron-positron pairs produced is large.

Two effects contribute to the increase in pair production due to
straggling.  Firstly, the value of the electron's $\eta$-parameter,
which controls the energy of the emitted gamma-ray, varies
considerably about its mean.  Occasional excursions to large $\eta$
result in the emission of higher energy gamma-rays with a
correspondingly larger probability of producing pairs.  Secondly,
straggling increases, on average, the time taken for electrons to
congregate in a region where few pairs can be created (a node with
$\bm{E}=0$ in our calculations), and this extends the time during
which high-energy photons are emitted and pairs produced.  We conclude
that straggling lowers the laser intensity required for significant
pair production.

In realistic situations, when large numbers of pairs are produced,
their mutual interactions as well as their back-reaction on the 
accelerating fields must be taken into account. In an
astrophysical setting, \cite{Timokhin} has addressed this problem by
combining a Monte-Carlo algorithm with a PIC-code.  We envisage that
the Monte Carlo algorithm we present will serve as a basis for a
coupled Monte Carlo-PIC algorithm aimed at modelling the initiation of
prolific pair-plasma creation in intense laser fields.

\ack We thank the UK Engineering and Physical Sciences Research Council for
support under grant number EP/G055165/1, entitled 'Multi-scale
simulation of intense laser plasma interactions'.

\clearpage

\appendix
\section{Estimation for the energy transferred to the laser field by a single particle}
\label{AppendixA}

From the momentum change expression at the photon emission time $ {\bf
  p}^{(\rm f)} = {\bf p} - \hbar {\bf k}$, we estimate the error, in the
Monte Carlo algorithm, related to the energy transferred to the laser
field by a single particle. The expression for the relative error
reads
\begin{eqnarray} \displaystyle\frac{\Delta \gamma}{\gamma} & \equiv &\frac{1}{\gamma} \left | \gamma^{(\rm f)}-\gamma+h\nu/(m_{\rm e}c^2) \right | \nonumber 
\end{eqnarray}
In the limit where the $1/\gamma$ and $1/\gamma^{(\rm f)}$ parameters are
small, it can be approximated as
\begin{eqnarray} \displaystyle\frac{\Delta \gamma}{\gamma} & \simeq & 
 \frac{1}{2} \frac{1}{\gamma} \left ( \frac{1}{\gamma^{(\rm f)}} -
 \frac{1}{\gamma} \right ) \quad  , \nonumber
\end{eqnarray}
which is of the order $1/(\gamma \gamma^{(\rm f)})$.


\section*{References}
\begin{thebibliography}{10}

\bibitem{ELI1}{Mourou G A, Labaune C L, Dunne M, Naumova N, and
  Tikhonchuk V T 2007} {Relativistic laser-matter interaction: from
  attosecond pulse generation to fast ignition} \textit{Plasma Physics
  and Controlled Fusion} 49 B667-B675

\bibitem{ELI2}{http://www.eli-laser.eu} {} \textit{}

\bibitem{Vulcan1}{Chekhov O, Collier J, Clark R J, Hernandez-Gomez C,
  Lyachev A, Matousek P, Musgrave I O, Neely D, Norreys P A, Ross I,
  Tang Y, Winstone T B, Wyborn B E 2009} {The 10 PW OPCPA Vulcan Laser
  Upgrade} \textit{IEEE} 978-1-4244-4080-1

\bibitem{Vulcan2}{ Hernandez-Gomez C, Blake S P,Chekhov O, Clarke R J,
  Dunne A M, Galimberti M, Hancock S, Holligan P, Lyachev A, Matousek
  P, Musgrave I O, Neely D, Norreys P A, Ross I, Tang Y, Winstone T B,
  Wyborn B E, and Collier J 2008-2009} {The Vulcan 10 PW project}
  \textit{CLF annual report, http://www.clf.rl.ac.uk}

\bibitem{salaminetal06}{Salamin Y I, Hu S X, Hatsagortsyan K Z, and
  Keitel C H 2006} {Relativistic high-power laser matter
  interactions}, \textit{Phys. Rep.} 427, 41-155

\bibitem{muelleretal09}{M{\"u}ller C, Hatsagortsyan K Z, Ruf M,
  M{\"u}ller S J, Hetzheim H G, Kohler M C and {Keitel} C H 2009}
  {Relativistic nonperturbative above-threshold phenomena in strong
    laser fields}, \textit{Laser Physics} 19, 1743-1752

\bibitem{dipiazzaetal09} {di Piazza A, Hatsagortsyan K Z and Keitel C
  H 2009} {Strong Signatures of Radiation Reaction below the
  Radiation-Dominated Regime}, \textit{Phys. Rev. Lett.} 102, 254802 

\bibitem{Bell}{ Bell A R and Kirk J G 2008} {Possibility of Prolific
  Pair production with High-Power Lasers}, \textit{Phys. Rev. Lett.}
  101 200403

\bibitem{PPCF}{Kirk J G, Bell A R and Arka I 2009} {Pair
  production in counter-propagating laser beams} \textit{Plasma
  Physics and Controlled Fusion} 51 085008

\bibitem{fedotov10}{Fedotov A M, Narozhny N B, Mourou G, and Korn G 2010} {Limitations on the attainable intensity of high power lasers} \textit{arXiv:1004.5398}

\bibitem{Sokolov1}{Sokolov I V, Naumova N M, Nees J A, Mourou G A,
  and Yanovsky V P 2009} {Dynamics of emitting electrons in strong
  laser fields} \textit{Phys. Plasma} 16 093115

\bibitem{Shen1}{Shen C S and White D 1972} {Energy Straggling and
  Radiation Reaction for Magnetic Bremsstrahlung} \textit{Physical
  Review Letters} 28 7

\bibitem{Shen2}{Shen C S 1972} {Magnetic Bremsstrahlung in an
  Intense Magnetic field} \textit{Physical Review D} 6 10

\bibitem{Erber}{ Erber T 1966} {High-energy electromagnetic
  conversion processes in intense magnetic fields} \textit{Rev. Mod. Phys.}
  38 626-59

\bibitem{SokolovQED}{Sokolov I V, Nees J A, Yanovsky V P, Naumova N
  M, Mourou G A 2010} {Emission and its back-reaction accompanying
  electron motion in relativistically strong and QED-strong pulsed
  laser fields} \textit{Phys. Rev. E} 81 036412

\bibitem{Timokhin}{Timokhin A N 2009} {Self-consistent
  modeling of pair cascades in the polar cap of a pulsar} \textit{Fermi
  Symposium, Washington D.C., Nov. 2-5}

\end{thebibliography}
\end{document}